\documentclass[conference,10pt]{IEEEtran}
\usepackage[pdftex]{graphicx}
\usepackage{url}
\usepackage{subfigure}

\usepackage{balance}

\usepackage{wrapfig}

\graphicspath{{00_grafx/}}
\usepackage{exscale,relsize}

\usepackage{booktabs}
\usepackage{multirow}

\newlength{\fcwidth}
\begin{document}

%\title{Voting for Rank-Authenticity in RPL}
%\title{Topology Authentication in RPL: Challenging the LIAR}
%\title{Topology Authentication: The LIAR within RPL}
%\title{Topology Authentication in RPL: The 13$^{th}$ Apostle}
\title{TRAIL: Topology Authentication in RPL}

\author{\IEEEauthorblockN{Heiner Perrey, Martin Landsmann, Osman Ugus}
%\IEEEauthorblockA{Hochschule f\"ur Angewandte\\Wissenschaften Hamburg\\
\IEEEauthorblockA{HAW Hamburg\\
firstname.lastname@haw-hamburg.de}
\and
\IEEEauthorblockN{Matthias W\"ahlisch}
\IEEEauthorblockA{Freie Universit\"at Berlin\\
m.waehlisch@fu-berlin.de}
\and
\IEEEauthorblockN{Thomas C. Schmidt}
%\IEEEauthorblockA{Hochschule f\"ur Angewandte\\Wissenschaften Hamburg\\
\IEEEauthorblockA{HAW Hamburg\\
t.schmidt@haw-hamburg.de
}
}

\maketitle
\begin{abstract}
%The IPv6 Routing Protocol for Low-Power and Lossy Networks (RPL) was recently introduced as the new routing standard for the Internet of Things (IoT). Although RPL defines basic security modes, it remains vulnerable to topological attacks which facilitate blackholing, interception, and resource exhaustion. We are concerned with analyzing the corresponding threats and protecting a future RPL deployment from topological attacks.

%The contributions of this paper are twofold. First, we analyze the state of the art, in particular the protective scheme VeRA, and present two new rank order attacks, as well as extensions of VeRA to mitigate them. Second, we derive and evaluate TRAIL, a generic scheme for topology authentication in RPL. TRAIL solely relies on the basic assumptions of RPL that (1) the root node serves as a trust anchor, and (2) each node interconnects to the root in a straight hierarchy. Using proper reachability tests, TRAIL can scalably and reliably identify any topological attacker without strong cryptographic efforts.
The IPv6 Routing Protocol for Low-Power and Lossy Networks (RPL) was recently introduced as the new routing standard for the Internet of Things. Although RPL defines basic security modes, it remains vulnerable to topological attacks which facilitate blackholing, interception, and resource exhaustion. We are concerned with analyzing the corresponding threats and protecting future RPL deployments from such attacks.

Our contributions are twofold. First, we analyze the state of the art, in particular the protective scheme VeRA and present two new rank order attacks as well as extensions to mitigate them. Second, we derive and evaluate TRAIL, a generic scheme for topology authentication in RPL. TRAIL solely relies on the basic assumptions of RPL that (1) the root node serves as a trust anchor and (2) each node interconnects to the root as part of a hierarchy. Using proper reachability tests, TRAIL scalably and reliably identifies any topological attacker without strong cryptographic efforts.

%\vspace{0.15cm}

%\textit{Keywords: IoT, routing security, mobile security, topology verification}

%This work presents two rank attacks which are not mitigated by VeRA. In the first attack, the adversary can decrease its rank arbitrarily. Hence, it can impersonate even the root node. In the second attack, the adversary can decrease its rank to the rank of any node within its access range. We present an enhancement for VeRA to mitigate the first attack. Additionally, a protocol for mitigating the second attack is introduced.
\end{abstract}

%\category{C.2.2}{Computer-Communication Networks}{Network Protocols}[Routing Protocols,Security]
%A category including the fourth, optional field follows...
%\category{D.2.8}{Software Engineering}{Metrics}[complexity measures, performance measures]

%\terms{Internet of Things, Security}


%\IEEEpeerreviewmaketitle
\section{Introduction}
\label{sec:intro}

%In this work we show threats and attack on the topology of RPL and present countermeasures to defend against these attacks.

%RPL \cite{RFC-6550} is a proposed standard by the \textit{Internet Engineering Task Force} (IETF) to provide an efficient and scalable routing protocol for low-power and lossy networks (LLN). RPL is feasible for energy restricted devices by minimizing the control plane traffic, which reduces the overall power consumption. This is especially beneficial when the devices are smallest sensor nodes, e.g. in (home) automation systems, smart-grids and surveillance systems.

RPL \cite{RFC-6550} has been designed as an efficient and scalable routing protocol for low-power and lossy networks (LLN). 
It promises to reduce the overall power consumption by minimizing the control traffic, which is a major requirement for the energy constrained devices envisioned in the future Internet of Things (IoT). Such tiny intercommunicating devices like sensor nodes used in (home) automation, smart grids or surveillance systems are expected to massively populate our environment soon.  
%It is feasible for energy restricted devices by minimizing the control plane traffic, which reduces the overall power consumption. This is especially beneficial when the devices are smallest sensor nodes, e.g. in (home) automation systems, smart-grids and surveillance systems.

RPL constructs one or several tree topologies oriented towards a single root node. Each node in the  RPL routing graph has a \textit{rank} derived from its parent relationship that describes the topological distance to the root.  Every node joining the topology calculates a higher rank than its parent, lower ranks are used for default upstream. This proactive organization leads to a Destination Oriented Directed Acyclic Graph (DODAG) topology, from which RPL is able to detect and remove inconsistencies reactively.

%The topology in RPL is hierarchically organized and oriented to the root node. Parent-child relations define the position of each node in the topology. The relationship between nodes is represented by the \textit{rank} of each node, which indicates the relative distance to the root. The rank increases monotonically, such that a node only selects parents with lesser rank value. The thus formed Destination Oriented Directed Acyclic Graph (DODAG) is created proactively. Inconsistencies are detected and removed reactively. 

Control traffic in this topology consists of \textit{DODAG Information Objects} (DIOs). A DIO advertises parameters and constraints for a specific DODAG that is uniquely identified by a version number. A node uses the information obtained from a DIO to select a parent node, compute its rank and join the DODAG, from which it inherits an upward route towards the root node. An optional upwards advertisement of \textit{Destination Advertisement Objects} (DAO) generates downward-oriented routes to children of a subtree. Depending on the mode of operation, these routes are either maintained and stored at each node (\textit{storing mode}), or forwarded to the root node and collected there (\textit{non-storing mode}).
%Downward routes are optional and maintained by the transmission of \textit{Destination Advertisement Objects} (DAO) to parent nodes. The mode of operation determines where the downward routing information is stored. Depending on the capabilities of the nodes, the information is either stored in each node (\textit{storing mode}) or only maintained by the root (\textit{non-storing mode}).
The integrity of distribution trees is essential for RPL, as an inconsistent hierarchy will lead to traffic redirections and a loss of routes to the root. In addition, RPL will attempt to cure tree deficiencies by reorganization, and a node that will hold up failures of the routing hierarchy may trigger repeated reconfigurations that drain resources of the network.

RPL offers basic protection against external attackers breaking into the topology \cite{RFC-6550}. However, as nodes may be captured and security keys can be extracted from them, the RPL topology is threatened by various attacks from inside the network \cite{msbcs-srdva-14}. The rank of a node and the DODAG version number are focal attributes in the topology. Known attacks are foremost based on them. A false  rank of a node forges the relative topological distance to the root and thus disarranges the hierarchy. An inconsistent version breaks the reference to the topological graph and causes the network to rebuild its routing graph. Corresponding protections are not part of the current RPL specifications.   

%Even though RPL offers basic protection against external attacks, is it still subject to various topology attacks, when faced with an internal attacker that has access to all security keys. Known attacks are mainly based on the rank and the version number. These attributes play a central role in RPL, as the the rank defines the position of a node in the topology and the version number allows the initiation of a global repair. An attacker closer to the root, has a greater impact on the network, as more traffic is forwarded to it. 

As major countermeasure, VeRA \cite{dhb-vnrar-11} has been proposed to fix these two classes of vulnerabilities by adding reverse hash chaining to DIO messages. Receivers shall be enabled to verify the advertised hierarchy. However, in the following we can show that VeRA remains vulnerable to rank attacks by forgery and replay. Furthermore, we present a more generic approach to solve the problem of topology authentication in RPL. Leaving aside the complexity of VeRA, our remaining work concentrates on a generic scheme for verifying RPL topologies. In detail, our contributions are the following:

\smallskip

\begin{enumerate}
  \item We analyze the incompleteness of message-rank-authentication in VeRA. 
  \item We present enhancements to VeRA for repair. 
  \item We introduce TRAIL (Trust Anchor Interconnection Loop), which can discover and isolate bogus nodes while these nodes attack the RPL routing hierarchy. TRAIL is derived of first hand principles and resolves the issues of topological infringements. 
  \item We implement TRAIL on the RIOT platform \cite{bhgws-rotoi-13} and evaluate it in a large-scale testbed.
  \item We will make our implementation openly available.
\end{enumerate}
\smallskip

%In this work we analyze the security approach VeRA \cite{dhb-vnrar-11}, which proposes a security scheme using hash chains to mitigate these attacks. Following up our previous work we identify topology attacks on RPL and vulnerabilities of VeRA. We propose adequate countermeasures as an enhancement for VeRA and  furthermore introduce our solution to defend against these threats in RPL. 
The remainder of this work is structured as follows. Section \ref{sec:rpl_security} discusses the problem of securing RPL, common attacks and related work. The incompleteness of VeRA is examined in Section \ref{sec:attack_vera}. Countermeasures for fixing VeRA are presented in Section \ref{sec:counter_vera}. Section  \ref{sec:liar} introduces and thoroughly evaluates TRAIL, our generic solution for topology authentication. Finally, we conclude in Section \ref{sec:conclusion} and look out on future work.

%RPL supports downward routes for a communication of the root with other nodes or a direct communication of nodes among each other. Downward routes are announced by \textit{Destination Advertisement Object} (DAO) messages, which are transmitted on upward routes. Downward routes are supported by two modes of operation, the \textit{storing} and \textit{non-storing mode}.

%In non-storing mode only the root stores source routes to all destinations. In storing mode a node stores and aggregates downward routing information provided by DAOs of its sub-DODAG. In turn the node announces an aggregated routing state in its DAO. Both modes mainly distinguish the way point-to-point communication is handled by the DODAG. In non-storing mode each message is forwarded to the root, which source routes the messages to its destination. In storing mode, all others nodes may route the message on a downward route.

%\input{01_sections/03_RelatedWork}
\section{RPL Security Challenges \& Related Work}
\label{sec:rpl_security}

\begin{figure}
 	\centering
 %	\twocolumn
 	\includegraphics[width=0.7\columnwidth]{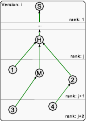}
 	\caption{\textsc{Regular RPL routing hierarchy}}
 	\label{fig:rpl_topology}
\end{figure}

RPL constructs a reverse path forwarding hierarchy by announcing tree parameters in the downward direction, starting from the root node. A node that successfully joined the tree advertises its rank towards its potential children in so called DIO messages, while unconnected nodes select as parent the neighbor of lowest rank, i.e., in closest position to the root. Following this algorithm, a fully connected acyclic, hierarchical graph is created in compliance to wireless reachability. Each of such DODAGs is associated with a unique version number to survey consistency. A regular RPL routing topology is displayed in Fig. \ref{fig:rpl_topology}.

%\iffalse
%\begin{wrapfigure}{l}{0.42\columnwidth}
%\end{wrapfigure}
%\par
%\fi

%The dynamic routing protocol RPL constructs a reverse path forwarding hierarchy by announcing tree parameters in downward direction, starting from the root node. A node that successfully joined the tree advertises its rank towards its potential children in so called DIO messages, while unconnected nodes select as parent the neighbor of lowest rank, i.e., in closest position to the root. Following this algorithm, a fully connected graph of unambiguous hierarchical relations is created in compliance to wireless reachability. Note that bidirectional links are required by RPL. Each of such DODAGs is associated with a unique version number to survey consistency. A regular RPL routing topology is displayed in Fig. \ref{fig:rpl_topology}. .

RPL specifies secured control plane messages for authenticity, integrity, and optional confidentiality \cite{RFC-6550}. Even though these basic security features defend against external attackers \cite{RFC-7416}, RPL remains unprotected against adversaries from inside the network \cite{dhb-vnrar-11,gk-rns-12}. Capturing a node and extracting security credentials enables an attacker to gain access to the control plane and to modify the routing topology. The rank and the version number are the key information for defining the structure of the routing system. The essential challenge for securing the routing topology thus is to protect rank and version number from any unwanted modification. Strong identity-based end-to-end authentification as introduced in \cite{msw-feaci-15} could defend a RPL routing system against internal modifications. However, its inherent complexity prevent this from being a standard solution. Next, we introduce the core attacks against the RPL topology and the assumptions made on the attacker.

%\begin{figure}%
%	\centering
%	\includegraphics[width=0.24\textwidth]{01_poster_rpl_graph}
%	%\caption{\textsc{Regular RPL routing hierarchy}}
%	\label{fig:rpl_topology}
%	\vspace{-2mm}
%\end{figure}%

%\begin{figure}	
%	%\subfigure[Regular RPL routing hierarchy]{\includegraphics[width=0.24\textwidth]{01_poster_rpl_graph}\label{fig:rpl_topology}}\qquad
%	\subfigure[Topology after a rank spoofing]{\includegraphics[width=0.24\textwidth]{01_poster_attack_forge}\label{fig:rpl_forge}}~%\qquad
%        \subfigure[Topology after a replay attack]{\includegraphics[width=0.24\textwidth]{01_poster_attack_replay}\label{fig:rpl_replay}}
%        \caption{RPL topology (b) with rank spoofing. The attacker $M$ propagates a rank $j_M$ falsely decreased by $\Delta$, and thereby incorrectly attracts nodes $1$, $2$, $4$, and the parent node $H$, which creates a sinkhole.  (c) visualizes a replay of the parent rank, only attracting nodes $1$, $2$, and $4$ with intact upstream to $H$.}
%    \label{fig:attack_rankforgery}
%\end{figure}

%We will now introduce the core attacks to the RPL topology along with a basic attacker model that minimizes assumptions.

\subsection{Attacker Model}
\label{sec:attacker_model}

We assume the presence of one or multiple attackers that physically captured and compromised multiple, arbitrary nodes on the network. The attacker has access to all available keys on the captured nodes, which include all information for joining and participating in the DODAG without restrictions. The compromised nodes are successfully integrated in the network and are thus authorized to transmit authenticated messages. Furthermore, the attacker is limited by the resources and constraints of the captured nodes. Hence, we assume that the attacker cannot install directed antennas or create multiple identities \cite{d-sa-02} to seemingly use several malicious nodes with one physical interface or to establish out-of-band channels. 
%Furthermore, the attacker is limited by the resources and constraints of the captured nodes. That is, the attacker cannot install directed antennas influencing the reception of multicast traffic, or create multiple identities \cite{d-sa-02} to seemingly use several malicious nodes with one physical interface.
The attacker aims at maximizing his impact on the network, for example by attracting as much traffic as possible for eavesdropping or sink-holing, or by affecting the operational conditions of as many nodes as possible.

%In our proposed counter approach in in section \ref{sec:liar} the assumption does not apply. Our approach forces the attacker to proof whether he keeps a connection to the root or not.
%For our own counter approach in section \ref{sec:liar}, we do not need to make this assumption, as we force the attacker to proof whether he keeps a connection to the root or not.

\subsection{Topology Attacks}
\label{sec:topology_attacks}
\label{sec:attacks_generell}

\subsubsection{Version Number Attacks}

The version number of the DODAG is increased by the root node, whenever a global repair is needed. This occurs, if inconsistencies cannot be repaired locally. In a version number attack \cite{dhb-vnrar-11}, an attacker illegally increases the version number of the DODAG. Publishing a higher version number will lead to a reconstruction of the RPL topology. This either serves as a preparation for a following attack such as on the rank, or can be repeatedly executed to disturb the network and drain the resources of nodes. 

%\begin{figure}	
%	%\subfigure[Regular RPL routing hierarchy]{\includegraphics[width=0.24\textwidth]{01_poster_rpl_graph}\label{fig:rpl_topology}}\qquad
%	\subfigure[]{\includegraphics[width=0.24\textwidth]{01_poster_attack_forge}\label{fig:rpl_forge}}~%\qquad
%        \subfigure[]{\includegraphics[width=0.24\textwidth]{01_poster_attack_replay}\label{fig:rpl_replay}}
%        \caption{RPL topology after a rank spoofing (a) and a replay attack (b).}
%    \label{fig:attack_rankforgery}
%%    \vspace{-5mm}
%\end{figure}
 
\begin{figure}	
	%\subfigure[Regular RPL routing hierarchy]{\includegraphics[width=0.24\textwidth]{01_poster_rpl_graph}\label{fig:rpl_topology}}\qquad
	\subfigure[Topology after a rank spoofing]{\includegraphics[width=0.24\textwidth]{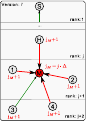}\label{fig:rpl_forge}}~%\qquad
        \subfigure[Topology after a replay attack]{\includegraphics[width=0.24\textwidth]{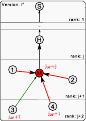}\label{fig:rpl_replay}}
        \caption{RPL topology (a) with rank spoofing. The attacker $M$ propagates a rank $j_M$ falsely decreased by $\Delta$, and thereby incorrectly attracts nodes $1$, $2$, $4$, and the parent node $H$, which creates a sinkhole.  (b) visualizes a replay of the parent rank, only attracting nodes $1$, $2$, and $4$ with intact upstream to $H$.}
    \label{fig:attack_rankforgery}
\end{figure}

\subsubsection{Rank Spoofing Attack}

In a rank spoofing attack \cite{dhb-vnrar-11}, a malicious node propagates an incorrect rank to change its position in the routing tree. Commonly, an attacker will choose a lower rank to improve its position in the hierarchy and achieve larger impact on the network. In response to forged rank advertisements, neighboring nodes select the attacker as parent and forward traffic towards it. Fig. \ref{fig:rpl_forge} visualises the topological manipulations caused by a strict rank decrease. The attacker $M$ propagates the lowest rank of the vicinity and attracts all its neighbors. In this example, the parent node $H$ is also attracted by the malicious node $M$, which creates a sinkhole. Node $3$ correctly selects $M$ as parent, but unknowingly propagates the illegal rank downtree. Thus, $3$ and its parents potentially attract even more children and increase the number of nodes that forward traffic towards the attacker $M$.

\iffalse
In general, there are two types of rank spoofing attacks. In the first type of attack, a malicious node chooses a better rank than all of its parents. In such a case, all traffic around the malicious node is forwarded to it and leads to a sinkhole within the network. In the second type of attack, the malicious node chooses two ranks. The first one is  correct and useful for staying connected with the root node. The second is a spoofed rank to become the parent of a large number of child nodes. If chosen not to go below the parent rank, the bogus rank advertisements will not interfere with the parent nodes.

When traffic is forwarded in RPL, nodes analyze each message regarding its rank with respect to the traffic direction. If the direction of a packet is inconsistent with the rank of its sender, an error-flag is set. To circumvent this consistency check, the attacker frauds both parents and children by seemingly maintaining consistent up- and downward routes. He uses a correct rank when communicating with parent nodes and forges only in downward direction. By modifying the rank properties, the attacker creates an inconsistency with respect to the RPL hierarchy. However, the internal inconsistency detection of RPL does not detect this.
\fi

\subsubsection{Rank Replay Attack}

An attacker who learned a valid rank from a (potential) parent may replay this value in its own advertisements and pretend to run at one hierarchy level above the proper value. This special case of a rank spoofing will not disconnect the attacker from the root as visualised in Fig. \ref{fig:rpl_replay}. In contrast to arbitrary rank forgery, the replay allows a malicious node to re-use a proper rank, even if rank verification schemes apply. We will show in the following section that present protection schemes are vulnerable to rank replay attack.

\subsection{Related Work}

Recent work has classified the different attacks on RPL \cite{mbc-tarit-16}, but only limited work has addressed the security of the RPL routing system. A security threat analysis for LLNs by the IETF \cite{RFC-7416} focuses on potential threats and attacks. However, the analysis solely proposes generic countermeasures to the described attacks. 
%The IETF security framework for RPL \cite{taddl-sfrlp-12} solely focus on protecting routing control against external attacks. 
Some attempts have been made to deal with topology attacks \cite{llyl-sbisr-11,dhb-vnrar-11, wp-esdtr-12, wrv-racri-13,msbcs-mtiar-15}. The authors in \cite{llyl-sbisr-11} propose an Intrusion Detection System to mitigate the rank attack and a local repair mechanism by installing additional monitoring nodes. The authors of \cite{msbcs-mtiar-15} present a mitigation strategy that allows nodes to dynamically adapt against a topological inconsistency attack based on the current network conditions. %\cite{wrv-racri-13}
VeRA addresses the rank and version number attacks by adding a rank and version control obtained from hash chaining \cite{dhb-vnrar-11,draft-dvir-roll-security-authentication}. 
%In detail, each DODAG version is represented by an element of a precomputed reverse hash chain $V_n,\ldots,V_0$, while for each version $i$ hash chained rank elements $R_{i,j}$ are computed from a one way hash function $h$ and random seeds $r$ and $x_i$ as follows
%
%\begin{equation}
%V_i  = h^{n+1-i}(r), ~~~~ R_{i,j} = h^{j+1}(x_i)
%\end{equation}
% 
%In the bootstrapping phase of a DODAG, the root disseminates the initial version hash element $V_{0}$, the actual version number VN$_{0}$, the last rank hash element $R_{1,l}$, a signature of the version hash, and the MAC of the last rank hash, $\langle V_0, \textrm{VN}_{0}, R_{1,l}, \{V_0, \textrm{MAC}_{V_1}(R_{1,l})\}_{sign} \rangle$. 
%
%Upon reception of a tree update message, a node will learn a new version element $V_i$ and the MAC of the current last rank hash $\textrm{MAC}_{V_i}(R_{i,l})$. It can then verify the new version by checking if $h^i(V_i) == V_0$ holds. Similarly, a new rank is verified by checking if $\textrm{MAC}_{V_i}(R_{i,l})~==~\textrm{MAC}_{V_i}(h^{l-j}(R_{i,j}))$ holds.  
While successfully mitigating a version number attack, the VeRA approach is still subject to two topology attacks \cite{lpuws-tar-13}. As we will work out in the following section, the first attack is a general \textit{rank spoofing}, which allows an attacker to pretend any rank and therefore any position in the DODAG. The second attack is a \textit{rank reply attack}, which allows an attacker to claim one level closer to the root by replaying its parent's rank. 
%In the second attack the adversary replays its parents rank, decrementing his rank by one.%, when replaying the rank of a parent in transmission range. 
Weekly and Pister \cite{wp-esdtr-12} concentrate on the evaluation of sinkhole attacks and their impact on the data throughput in RPL. They utilize a rank authentication based on VeRA and introduce a parent fail-over technique to blacklist sinkhole nodes. The root maintains a list of nodes that go below a threshold, which defines the minimum expected data receptions for each node. A node that finds itself on the list, blacklists its default next-hop towards the root, since some node on the path seems to not forward traffic. %They evaluate the data throughput using these security techniques. 
The authors observe that an adversary can attack VeRA by replaying the rank of his parent. Similarly, Wallgren et al. \cite{wrv-racri-13} propose to maintain a whitelist in combination with a heartbeat protocol in which the root periodically sends echo requests to each node to check the connectivity. A node that does not respond is considered malicious and is thus removed from the whitelist. 

Our work goes beyond mitigating sinkhole attacks. By performing generic topological tests, we inquire on the integrity of the routing hierarchy, identify and isolate individual attackers. The rank protection approaches discussed above rely on the authentication of digital signatures. Authentication in RPL is implemented either by an asymmetric signature scheme like RSA \cite{rsa-modsp-78}, or a symmetric message authentication code like CBC-MAC \cite{RFC-3610} with a pre-shared key. Even though the complex use of asymmetric cryptography in wireless LLNs remains a challenge for class 1 devices \cite{wgegs-eapkc-05, plp-pkciw-06, gpwes-ceccr-04}, it adds the advantage to unambiguously authenticate the sender of a message \cite{msw-feaci-15}. Conversely, when a MAC is created from a group key, it suffers the disadvantage of authenticating {\em any} sender from that group. Recent work on pre-computation techniques \cite{abcp-lssws-13} strengthened the case for standard signature deployment on sensor nodes.  In this work, we rely on the RPL root node as a trust anchor, since it is commonly deployed as a more powerful gateway node. Signature creation remains bound to this RPL root. 
 % \cite{plp-pkciw-06, gpwes-ceccr-04
\begin{table}

	\caption{Glossary of Notations}
	\label{tab:nomen}
	\begin{center}

%\tiny
	\begin{tabular}{ c l }
        \toprule
		%$\textsc{Symbol}$ & $\textsc{Definition}$  \\% = $enc_{k_i}()$}\\
		Symbol & Definition  \\% = $enc_{k_i}()$}\\
		\midrule
		$i$ &	 Index for Version Hash Chain ($0,\dots,n$)\\
		$j$ &	 Index for Rank Hash Chain ($0,\dots,l$)\\
		$l$ &	 Last Index of Rank Hash Element ($= 2^{16}-1$) \\ %l+1 = depth of rank hash chain
		$n$ & Last Index of Version Hash Element \\
		$V_{i}$ & $i$-th Element of Version Hash Chain\\
		$R_{i, j}$ & $j$-th Element of Rank Hash Chain for $i$-th Version\\
		$r$ & Random Seed for Version Hash Chain\\
		$x_{i}$ & Random Seed for Rank Hash Chain at Version $i$\\ 
		VN$_{i}$ & Numeric Version Number at Version $i$ \\
%		$h(\cdot)$ & One-Way Hash Function\\
%		$enc_k(\cdot)$ & Symmetric Encryption of $\cdot$ with Key $k$\\
%		$dec_k(\cdot)$ & Symmetric Decryption of $\cdot$ with Key $k$\\
		$c_{i}$ & $i$-th Element of Encryption Chain\\
%		$\DELTA$ & Rank-Steps of Rank-Forgery \\
%		$REQ$ & Rank Validation Token Request Message\\
%		$RESP$ & Rank Validation Token Response Message\\
%		$ID$ & Node Identifier, e.g. MAC- or IP-Address\\
%		$REQ$ & Rank Validation Token Request Message\\	
				
        \bottomrule
	\end{tabular}
	%	\vspace{1mm}\\
		%\normalsize
\end{center}
%\vspace{-5mm}
\end{table}

\section{Attacking VeRA}
\label{sec:attack_vera}

%\subsubsection{Protocol Overview}

%We describe the VeRA protocol and evaluate its security.

\subsection{VeRA in a Nutshell}
\label{subsec:vera}
 VeRA is performed in two steps: initialization and version number update. The scheme assumes that the nodes are given a public key $pk$ for a public key signature scheme like RSA, and the corresponding secret key $sk$ is known only to the DODAG root.

\subsubsection{Initialization}
%{\em Initialization:}
The DODAG root generates the hash chains to be used for securing version number and rank updates. For $n$ version updates, the root picks a random number $r$, a secure hash function $h$,  and computes a hash chain, $\{r,V_i\}_{i=0,\cdots, n}$, with  $V_i = h^{n+1-i}(r)$. Additionally, the root generates a rank hash chain of size $l+1$ for each version $V_i$. Let $R_{i,0}, \cdots, R_{i,l}$ denote the rank hash chain for $V_i$. Then, its elements are computed as $\{R_{i,j} = h^{j+1}(x_i)\}_{j=0,\cdots,l}$, where $x_i$ is a random number. Subsequently, for bootstrapping the security, the root broadcast an initialization message $\{V_0,{VN}_0,MAC_{V_1} (R_{1,l}), \sigma\}$ to all nodes in a DIO message. Thereby, ${VN}_0$ denotes an initial version number chosen by the root and $\sigma =Sig_{sk}(V_0,{VN}_0,MAC_{V_1} (R_{1,l}))$ the signature. Each node stores this message after verifying the signature using the public key $pk$.  

\subsubsection{Version number update} 
%{\em Version number update:}
To update the version of a DODAG from ${VN}_{i-1}$ to ${VN}_{i}$, the root sends a DIO message $\{{VN}_{i}, V_i, MAC_{V_{i+1}}(R_{i+1,l}), R_{i,Rank_{sender}}\}$, where $Rank_{sender}$ is its new rank\footnote{For the root, $ R_{i,Rank_{sender}} = h^{Rank_{root}}(x_i)$.}. Each intermediate node receiving this message checks first whether the new version number is higher than the current one, i.e., if ${VN}_{i} > {VN}_{i-1}$. If this is the case, it continues to verify that the version update was indeed initialized by the root by checking if $V_0 = h^{({VN}_{i} - {VN}_{0})} (V_i) = h^i(V_i)$ holds. If any of these verifications fails, the node terminates the version update operation. Otherwise, it proceeds with verifying the rank  of its parent by checking the hash chain consistency, i.e., $MAC_{V_i} (h^{l-Rank_{parent}}(R_{i,Rank_{parent}}))~=~MAC_{V_{i}}(R_{i,l})$.  Note that $MAC_{V_{i}}(R_{i,l})$ was received in the previous update, while $V_i$ is received in the current update.  Finally, the child node calculates its own rank $\tau$ using the objective function and forwards the received DIO message to nodes lower in the topology with the corresponding rank chain element $ R_{i,Rank_{sender}} = h^{(\tau - Rank_{parent})} ( R_{i,Rank_{parent}})$.

\subsection{(In)Security of VeRA}

The security of VeRA relies on the assumption that increasing the version number or decreasing the rank value requires an attacker to compute the pre-image of a hash chain element. However, due to the stateful nature of the VeRA protocol, the pre-image resistance of the hash chains alone are not sufficient for security.  VeRA is a stateful protocol, since the security of each version update relies on the parameters revealed in a previous update. Although, the initialization message is signed, as shown in Section~\ref{sec:attack_impersonation}, it is not sufficient to mitigate rank chain forgery performed by malicious insiders or when jamming attacks are considered. Hence, additional methods preserving backward secrecy of the rank hash chains are needed to mitigate VeRA against such attacks\footnote{Solutions based on time synchronization are not considered in this work.}. Within the scope of the VeRA protocol, we give the following definitions:

%Consider the version hash chain elements $\{r,V_i\}_{i=n,\cdots,0}$, the corresponding rank hash chains $\{x_i, R_{i,l}= h^{l+1} (x_i)\}_{i=n,\cdots,1}$, and the signature message $\sigma$ described above. 
%Within the scope of the VeRA protocol, we give the following definitions:

\smallskip

\newtheorem{definition}{Definition}
\begin{definition}[Perfect-backward-secure version update protocol]
%\textbf{Definition} 
%\emph{Perfect-backward-secure version update protocol:}
A version update protocol is perfect-backward-secure if an adversary cannot efficiently calculate a valid rank hash chain $\{x'_i,R'_{i,l}\}_{i \in \{n,\cdots,1\}}$ with $x_i \neq x_i'$ even if it is given all elements of the version hash chain, i.e., $V_n$\footnote{All other elements $V_{n-1},\cdots, V_0$, can be calculated from $V_{n}$}, the corresponding rank hash chains $\{x_i, R_{i,l}\}_{i=n,\cdots,1}$, and the signature $\sigma$. A hash chain $\{x'_i,R'_{i,l}\}$ is valid, if its verification in the $i$th version update, i.e., $\{{VN}_{i}\}_{i \in \{n,\cdots,1\}}$, at any receiving node returns a success.
\end{definition}

\smallskip

\begin{definition}[$\lambda$-backward-secure version update protocol]
%\textbf{Definition} 
%\emph{$\lambda$-backward-secure version update protocol:}
A version update protocol is $\lambda$-backward-secure if an adversary cannot efficiently calculate a valid rank hash chain $\{x'_i,R'_{i,l}\}_{i \in \{n,\cdots,1\}}$ with $x_i \neq x_i'$ even if it is given up to $\lambda < n$ elements of the version hash chain, i.e., $V_\lambda$, the corresponding rank hash chains $\{x_i, R_{i,l}\}_{i=\lambda,\cdots,1}$, and the signature $\sigma$. A hash chain $\{x'_i,R'_{i,l}\}$ is valid, if its verification in the $i$th version update, i.e., $\{{VN}_{i}\}_{i \in \{n,\cdots,1\}}$, at any receiving node returns a success.
\end{definition}

%\begin{lemma}
\noindent \textit{Lemma.} The VeRA protocol is a $\lambda$-backward-secure version update protocol.
%\end{lemma}

\smallskip

\begin{IEEEproof}[Proof (Sketch)]
Consider a VeRA setting for $n=3$. Furthermore, consider the version hash chain $\{r,V_3,V_2,V_1,V_0\}$, and the rank hash chains $\{\{x_3, R_{3,l}\},\{x_2, R_{2,l}\},\{x_1, R_{1,l}\}\}$. Given the version hash chain element $V_2$, the adversary can calculate a valid hash chain  $\{x'_2,R'_{2,l}\}$ by simply picking a random number $x_2' \neq x_2$ and subsequently authenticating $R'_{2,l}$ with the MAC using $V_2$ as the key. This rank hash chain would be verified as valid in a version update $V_2$. Hence, the VeRA protocol is $2$-backward-secure version update protocol. That is, it remains secure against rank hash chain forgery as long as no version hash chain element $V_{\lambda\ge 2}$ is compromised.   
\end{IEEEproof}
%\paragraph{Praxis relevance of backward-secrecy}
%\begin{remark}

\textit{Remark} {\em Praxis relevance of backward-secrecy.}
RPL is  a routing protocol for LLNs. A typical characteristic of such networks, such as WSNs, is that they are often deployed in public and even in hostile environments. Hence, they are typically easy to access by attackers. Wireless communication used in such networks allows an attacker to disrupt and even entirely block the communication between nodes. For instance, (selective) jamming attacks \cite{xmtz-jsnad-06} allow to partition a network. Similarly, selective-forwarding attacks \cite{kw-srwsn-03} allow the attacker to drop selected packets during routing. Such attacks allow the adversary for decreasing its own rank and, hence, the rank of those nodes located in its sub-DODAG if the version update protocol used is only  $\lambda$-backward-secure like VeRA. In the following, we describe a practical rank chain forgery attack in  existence of e.g., selective-forwarding or jamming attacks.
%\end{remark}

%In VeRA,  the version number, each rank, and their corresponding hash elements are linked to the initial signature disseminated in the bootstrap message. Hence, an attacker is not able to increase the version number, or arbitrarily change its rank without further effort. However, the tie between the hash chains and the signature breaks after the first version update, as both hash chains evolve independently. 

%\subsection{Update Withholding  Attack}
%\paragraph{Update Withholding  Attack}
%\paragraph{Rank hash chain forgery attack}
\subsubsection{Rank hash chain forgery attack}
\label{sec:attack_impersonation}

VeRA is only $2$-backward-secure. Hence, it is vulnerable to rank chain forgery. Such an attack might be performed as follows. The DIO messages for two subsequent version updates ${VN}_{i}$ and ${VN}_{i+1}$ are prevented from being received by all or some of the nodes within the network. This can be achieved e.g., through a selective-forwarding  or (selective-)jamming attack on the DIO messages of the version updates. After receiving the  hash chain element $V_{i+1}$ in the version update ${VN}_{i+1}$, the attacker calculates a bogus hash chain  $\{x'_{i+1},R'_{i+1,l}\}$ by simply picking a random number $x'_{i+1}$ and subsequently authenticating it with the MAC using $V_{i+1}$ as the key. Subsequently, the blocked version update ${VN}_{i}$ is resumed by forwarding the DIO message containing the MAC of the forged rank hash chain $MAC_{V_{i+1}}(R'_{i+1,l})$. Finally, once the version update ${VN}_{i}$ is completed, the version update ${VN}_{i+1}$ is initiated, in which the attacker can claim an arbitrary rank value.

%A node receiving advertisements from the attacker cannot distinguish whether the hash chain has been forged or created truthfully, as the information received is consistent and verifiable. 
%Still, if the attacker communicates with his parents, he may have to use his \textit{correct} rank, as RPL would detect rank inconsistencies otherwise (see Section \ref{sec:attacks_generell}).

%\subsection{Rank Replay Attack}
%\paragraph{Rank Replay Attack}
\subsubsection{Rank replay attack}
\label{sec:attack_replay}

%The rank replay attack introduced in Section \ref{sec:topology_attacks} applies to VeRA. 
In each rank update, VeRA discloses the cryptographic credentials needed for verifying the advertisements from parents to each node. These credentials are not bound to any sender-specific attributes. Hence, a malicious node can transparently forward them down the tree to decrease its rank. As visualized in Fig.  \ref{fig:rpl_replay},  a malicious node $M$ receives valid rank announcements from the honest node $H$. This includes the version hash $V_i$, its rank $j$, and the associated hash element $R_{i,j}$. It can simply re-use them in its own rank advertisements to nodes $1 \ldots 4$ for gaining one hierarchy level.   In consequence, the honest nodes $1$, $2$, and $4$ can verify the bogus rank announcements and prefer $M$  over $H$  due to  better connectivity.  Node $3$ correctly selects $M$ as parent, but calculates a falsely improved rank. All children of $M$ propagate maliciously lowered ranks down the sub-DODAG.

\section{Fixing VeRA}
\label{sec:counter_vera}

We introduce two countermeasures to fix  VeRA against the described attacks. Our first countermeasure makes the VeRA approach perfect-backward-secure and, hence, it mitigates the rank hash chain forgery attacks.  Our second countermeasure is a simple challenge-response procedure proposed for mitigating the rank replay attacks.  

%In this section, we propose countermeasures to overcome the withholding and the replay attacks on VeRA. We describe (a) a recursive trust establishment scheme which mitigates the update withholding attack, and (b) a protocol extension based on a challenge-response procedure which mitigates the rank replay attack. Note that these countermeasure can also be applied to other authentication approaches that are based on hash chains.

%We design (a) a recursive trust establishment scheme which correlates the version number \emph{and} the rank, and (b) a protocol extension based on a challenge-response procedure. Note that these countermeasure can also be applied to other authentication approaches that are based on hash chains.

%In this section we propose countermeasures for the attacks against VeRA summarized in section \ref{sec:attack_vera}. We show that using an encryption chain provides the necessary correlation between the version number and rank hash chain for a recursive trust establishment. To mitigate the rank-replay attack, we introduce an additional enhancement for VeRA based on a challenge-response procedure.
%Additionally, we solve the remaining threat of a rank-replay attack by introducing a protocol enhancement based on a challenge-response procedure.

\subsection{VeRA++: Perfect-backward-secure VeRA}
VeRA authenticates a rank hash chain for a version $V_i$ using a MAC keyed with $V_i$. In each version number update a MAC key is revealed. Hence, VeRA provides only the $\lambda$-backward-secrecy. To achieve the perfect-backward-secrecy, we propose to authenticate the rank hash chains using an encryption chain instead of MACs. In the following, we first describe the construction of the proposed encryption chain. Subsequently, we describe the VeRA++ approach, i.e. an extension of VeRA with the proposed encryption chain. Finally, we show that VeRA++ provides the perfect-backward-secrecy.

%\paragraph{Construction of the encryption chain}
\subsubsection{Construction of the encryption chain} 
After generating the version number hash chain and the rank hash chains $\{x_i,R_{i,l}\}_{i=n,\ldots,1}$ as described in Section~\ref{subsec:vera}, the root node computes the (rank) encryption chain $\{c_i\}_{i=n,\dots,1}$ as follows: $c_n$ is set to the last element of the rank hash chain for $V_n$, i.e., $c_n = R_{n,l}$. Subsequent elements of the encryption chain $c_i$ are calculated by encrypting the last element of the corresponding rank hash chain $R_{i,l}$ using $c_{i+1}$ as the encryption key. That is, $\{c_i=enc_{c_{i+1}}(R_{i,l})\}_{i=n-1,\ldots,1}$, where $enc$ is a symmetric key encryption scheme such as AES.
%For $i=n-1,\cdots,1$, each element of the encryption chain $c_i$ is calculated by encrypting the last element of the corresponding rank hash chain $R_{i,l}$ using $c_{i+1}$ as the encryption key. That is, $\{c_i=enc_{c_{i+1}}(R_{i,l})\}_{i=n-1,\ldots,1}$, where $enc$ is a symmetric key encryption scheme such as AES.
%\begin{equation}
%  \{c_i=enc_{c_{i+1}}(R_{i,l})\}_{i=n-1,\ldots,1}, 
%\end{equation} where
%$enc$ is a symmetric key encryption scheme such as AES.

\subsubsection{Extension of VeRA with the encryption chain}

In VeRA++, the initialization and version number update steps are performed slightly different than in VeRA. In the initialization step, the root broadcast the initialization message $\{V_0,{VN}_0,c_1, c_n, \sigma\}$ to all nodes in a DIO message. Thereby, $\sigma =Sig_{sk}(V_0,{VN}_0,c_1, c_n)$. As in the VeRA, each node stores this message after verifying the signature with  $pk$. 

In the version number update step, to update the version of DODAG from ${VN}_{i-1}$ to ${VN}_{i}$, the root sends a DIO message $\{{VN}_{i}, V_i, c_{i}, R_{i,Rank_{sender}}\}$. Similar to the VeRA, each intermediate node receiving first checks if the new version number is higher than the current one, i.e., if ${VN}_{i} > {VN}_{i-1}$. If this is the case, it continues to verify that the version update was indeed initialized by the root by checking if $V_0 = h^{({VN}_{i} - {VN}_{0})} (V_i) = h^i(V_i)$ holds. If one of these verifications fail, the node terminates the version update operation. Otherwise, it proceeds with verifying the rank  of its parent.

Assume that the parent node claims to have the rank value $Rank_{parent}$. The child node verifies its validity by checking if $h^{l-Rank_{parent}} = dec_{c_{i}}(c_{i-1})$. Note that $c_{i-1}$ was received in the previous update. A successful verification implies that the rank of its parent is increasing monotonically\footnote{In the last version update, no decryption is required since $c_n = R_{n,l}$.}. Subsequently, the child node calculates its own rank $\tau$ using the objective function and forwards the received DIO message to the nodes lower in the topology as in VeRA. 
%lower-rank nodes with a corresponding rank chain element, i.e., $ R_{i,Rank_{sender}} = h^{(\tau - Rank_{parent})} ( R_{i,Rank_{parent}})$. 

\subsubsection{Security of VeRA++}
We show that the VeRA++ approach is a perfect-backward-secure version update protocol.

%\begin{proposition}
\noindent
\textit{Proposition.}
The VeRA++ approach described above is a perfect-backward-secure version update protocol if the  underlying encryption function $enc$, the signature scheme $Sig$, and the hash function $h$ are cryptographically secure.
%The VeRA++ approach described above is a perfect-backward-secure version update protocol if the  underlying encryption function $enc$ and the signature scheme $Sig$ are secure and the underlying hash function $h$ is pre-image resistant.
%\end{proposition}

\smallskip

\begin{IEEEproof}[Proof (Sketch)]
Assume that the VeRA++ approach  a $\lambda$-backward-secure version update protocol. Then, according to our definition, given a hash chain element $V_\lambda$ with $\lambda <n$ and the encryption chain $\{c_i\}_{i=\lambda,\cdots,1}$ and the rank hash chains $\{x_i, R_{i,l}\}_{i=\lambda,\cdots,1}$ and the signature $\sigma$, there exist an efficient algorithm that calculates a forged rank hash chain $\{x'_i, R'_{i,l}\}$ with  $x_i \neq x_i'$ which is valid for a version update $\{{VN}_{i}\}_{i \in \{n,\cdots,1\}}$.

\begin{itemize}
\item
%$\bullet$ 
$i=1$: $c_1$ is signed. Thus, there is only one possibility for calculating a forged rank hash chain $\{x'_1, R'_{1,l}\}$  with  $x'_1 \neq x_1$. The adversary needs to find a $c'_2$ such that $R'_{1,l} = dec_{c'_2}(c_1)$ for an arbitrarily chosen $x'_1$. The probability of finding such inputs ($c'_2, c_1, x'_1$) is negligible if the underlying encryption function $enc$ is secure and the hash function $h$ is pre-image resistant and the signature scheme is secure.
\item
%$\bullet$ 
$i=\lambda$: For calculating a forged rank hash chain $\{x'_\lambda, R'_{\lambda,l}\}$ with  $x'_\lambda \neq x_\lambda$, the adversary needs to find a $c'_{\lambda+1}$ such that $R'_{\lambda,l} = dec_{c'_{\lambda+1}}(c_\lambda)$ for an arbitrarily chosen $x'_\lambda$. However, the  probability of such inputs ($c'_{\lambda+1}, c_\lambda, x'_\lambda$) is negligible if the underlying encryption function $enc$ is secure  and the hash function $h$ is pre-image resistant and the signature scheme is secure\footnote{If the adversary would be allowed to choose both $c'_{\lambda+1}$ and $c'_\lambda$, it could easily calculate $R'_{\lambda,l}$. However, choosing $c'_\lambda$ is not allowed: each node accepts the $c_{\lambda+1}$ only if $dec_{c_{\lambda+1}}(c_\lambda)$ yields a valid rank hash chain $R_{\lambda,l}$, where $c_1$, i.e., $R_{1,l}$ is signed. Thus, for choosing a new $c'_\lambda \neq c_\lambda$, the adversary needs to forge the signature $\sigma$. %since starting from $c_1$, each node accepts the next element of the encryption chain $c_{\lambda+1}$ only if $dec_{c_{\lambda+1}}(c_\lambda)$ yields a valid rank hash chain $R_{\lambda,l}$, where $c_1$, i.e., $R_{1,l}$ is signed. Thus, to choose $c'_\lambda$, the adversary needs to forge the signature $\sigma$.
}.
\item
%$\bullet$ 
$i=n$:  $c_n = R_{n,l}$ is signed. Thus, there is only one possibility for calculating a forged rank hash chain$\{x'_n, R'_{n,l}\}$ with  $x_n \neq x_n'$. The adversary needs to forge the signature message $\sigma$. However, the probability of a signature forgery is negligible if the underlying signature scheme is secure. 
\end{itemize}

This shows that a rank hash chain forgery for $\lambda \leq n$ requires to break the security of either $h$ or $enc$ or $Sig$. Hence, given the secure instances of algorithms $h$, $enc$, and $Sig$, according to our definition, the VeRA++ protocol is  a perfect-backward-secure version update protocol. 
\end{IEEEproof}

\subsection{Countermeasure against Rank Replay Attacks}

To mitigate rank replay attacks, we introduce a challenge-response  scheme bASED on the rank hierarchy implemented by RPL. A malicious node, claiming a lower rank value than its actual  value, is challanged to prove that it has a parent node of lower rank than the claimed.

%To mitigate rank replay attacks, we introduce a challenge-response based scheme. This scheme incorporates intrinsic properties of the RPL topology. A suspicious node needs to prove a relation with a valid parent of its currently claimed rank.

\subsubsection{General Idea} %The general idea is based on the following observation. 
Each RPL node receives the rank hash chain element of its parent to verify the parent's rank as well as to calculate its own rank.  A parent node, claiming to have the rank $j-1$ in a version update $i$, sends the hash chain element $R_{i,j-1}$ to its children. Each child receiving this message verifies it by checking if $R_{i,l}= h^{l-(j-1)}(R_{i,j-1})$ holds. However, due to pre-image resistance of the hash function $h$, they cannot calculate $R_{i,j-2}$, i.e., the hash chain element valid for their grandparent nodes.  Our scheme relies on this one-way property of the rank hash chains. That is, any node claiming to have a rank $j$ knows (or can calculate) the hash chain element of their parents, their own, and their children only, i.e. $\{R_{i,k}\}_{k = j-1,\ldots,l}$. The hash chain elements for lower ranks (e.g.., for their grandparents) $\{R_{i,k'}\}_{k' = 1,\ldots,j-2}$ remain unknown to them (or cannot be calculated by them).  Hence, any node claiming to have a rank $j$ must be able to encrypt a challenge message using $R_{i,j-1}$ as the key, correctly. An attacker that incorrectly replays the rank hash element of its own parent cannot encrypt such a challenge, since it needs to know the key $R_{i,j-2}$ in such a case.

%Any RPL node knows the rank hash elements distributed by its parents and can construct the hash elements of its children. However, it does not know (and cannot construct) the rank hash elements of its grandparent nodes. Claiming rank $j$ correctly, implies that the node knows $R_{i,j}$ and $R_{i,j-1}$, but it remains unaware of the upper rank hash elements $R_{i,j'}$ with $j' < j-1$. Consequently, a challenge that asks for encryption with $R_{i,j-1}$ can be solved by all nodes with rank $\leq j$. An attacker that incorrectly replays the rank hash element of its own parent cannot solve such a challenge.

%Any RPL node knows the rank hash elements of its direct parents. A rank hash element $R_{i,j}$ thus describes a secret between all nodes from the root up to all nodes at rank $j$. 

\subsubsection{Challenge-Response Scheme} 
\begin{figure}	
	\subfigure[Replaying rank announcement]{\includegraphics[width=0.24\textwidth]{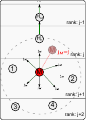}\label{fig:replay_prevention_init}}~
	\subfigure[Failing the challenge-response]{\includegraphics[width=0.24\textwidth]{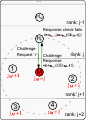}\label{fig:replay_prevention_challenge_response}}\qquad
        \caption{(a) The attacker $M$ replays the rank announcement via multicast and creates a malformed local topology. (b) Its parent $H_P$ is in transmission range and receives the replay, as well. The honest node challenges the attacker by sending a random number $r$. The challenge can only be solved if $M$ has a relation with the grandparent node $H_{GP}$, i.e., knows $R_{i,j-1}$.}
    \label{fig:replay_prevention}
\end{figure}

Assume that the node $M$ depicted in Fig.~\ref{fig:replay_prevention} is suspicious of replaying the rank $j$ of its parent $H$ to obtain an improved position in the DODAG topology. To verify the rank of $M$, the parent node $H$ constructs a challenge message $\langle \textnormal{ID}_M, r \rangle$, which includes the host ID of $M$ ID$_M$ and a random number $r$. This message is to be encrypted by $M$ with $R_{i,j-1}$. $M$ shall reply with $enc_{R_{i,j-1}}(\langle \textnormal{ID}_M, r \rangle)$ to $H$. $H$ can then check whether $M$ holds the correct rank. If $M$ cannot solve the challenge, it has no valid parent of the claimed rank and incorrectly announced $j$.

%Assume node $M$ is suspicious of replaying the rank $j$ of its parent $H$ to obtain an improved position in the DODAG as shown in Fig.~\ref{fig:replay_prevention}. Parent $H$ constructs a challenge message $\langle \textnormal{ID}_M, r \rangle$, which includes the host ID of $M$ ID$_M$ and a random number $r$. The message is to be encrypted by $M$ with $R_{i,j-1}$. $M$ shall reply with $enc_{R_{i,j-1}}(\langle \textnormal{ID}_M, r \rangle)$ to $H$. $H$ can then check whether $M$ holds the correct rank. If $M$ cannot solve the challenge, it has no valid parent of the claimed rank and incorrectly announced $j$.

\subsubsection{Applying the Challenge-Response Scheme}

The challenge-response scheme is initiated by a RPL node that complies with two requirements: (a) it is at the same rank level as claimed by the attacker, (b) it is within the transmission range of the attacker. The first requirement allows for self-organization among RPL nodes according to correct and incorrect ranks (i.e., who initiates the challenge). The second requirement is necessary to react on  suspicious rank announcements (i.e., observing rank upgrade). It is noteworthy that these requirements fully comply to the RPL protocol message design.

%After an RPL node $H$ identifies an attacker $M$, it needs to exclude the attacker from the network. $H$ can initiate a local repair, which starts a complete rebuilding of the sub-DODAG. $H$ can also inform the root about $M$. The root, which represents the trust anchor of the complete DODAG, informs the children of $M$ about the incorrect rank.

Our approach detects the potential attack by requiring each node to multicast its rank to all neighbors. If a parent $H$ detects an inconsistent routing state, which is not removed by a local repair, the suspicious node is challenged to proof its rank. If the node does not pass the challenge, either the sub-DODAG can simply be excluded from upward routing, or the root node can be included in the validation process. The root creates a validation packet for the children of the malicious node $M$. Finally, as many surrounding nodes as possible are informed about the rank state of node $M$. This gives each node the ability to independently react and for example discard $M$ as a routing node. Note that the root node can trust $H$ after the update is delivered successfully and the replay detection is applied recursively from $H$ to the root.

\smallskip
%\begin{remark}
\noindent
\textit{Remark.} Our challenge-response scheme is not secure in existence of out-of-band channels or $k$ directly connected attackers. An attacker, who can obtain the rank hash chain element for a rank $j-\Delta$ through such a tunnel, can correctly respond any challenge issued for verifying the ranks $\leq j-\Delta$. Another limitation of this approach is that the children of a malicious cannot be reliably notified about an attack. Introducing a white- or blacklist containing benign or malicious nodes respectively, as proposed in \cite{wrv-racri-13} and \cite{wp-esdtr-12}, holds the drawback of straining the entire network with local information. Hence, in the next section we propose TRAIL which inverses the direction of rank validations.

\section{TRAIL -- Trust Anchor Interconnection Loop}
\label{sec:liar}
We introduce \textit{TRAIL}, our generic approach to detect and prevent topological inconsistencies. In contrast to the previous approaches, each node is enabled to validate its upward path to the root and to detect rank spoofing on it. Our test furthermore identifies the largest sub-DODAG(s) affected by non-monotonous rank order. Having learned such inconsistency, the root of that sub-DODAG may either trigger a local repair, or disconnect its malicious sub-tree and rely on alternate paths.
In the following, we treat ranks as monotonously increasing integers. It is noteworthy that any RPL rank function is monotonous and can be reverted to an integer chain.

\subsection{TRAIL Idea: Path Validation}

\begin{figure}	
	\subfigure[Successful Rank Validation]{\includegraphics[width=0.24\textwidth]{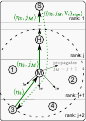}\label{fig:trail_single}}~
	\subfigure[Failed Rank Validation]{\includegraphics[width=0.24\textwidth]{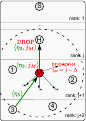}\label{fig:trail_single_attack}}\qquad
        \caption{\textsc{TRAIL Single Rank Validation}: Node $3$ initiates the rank validation by sending a nonce, $\eta$, to parent $M$. In (a) $M$ announces its true rank. The message arrives at the root and is singed. In (b) $M$ uses a forged rank, hence the message is dropped.}
    \label{fig:trail_single_validation}
\end{figure}

The key idea of TRAIL is to validate upward paths to the root using a round trip message.  Without relying on encryption chains as in VeRA(++), a node can conclude rank integrity from a recursively intact upward path.  

A child node that received a rank advertisement from its parent initiates a {\em positive attestation} of the rank as follows. It sends a test message with a random nonce $\eta$ upwards to its parent. The parent adds its rank $j$ and forwards the test message  $\langle j, \eta \rangle$ upstream towards the root.  At each intermediate hop, the receiving upper node verifies that (a) the rank in the test message is higher than its own, and (b) the rank of the sending node lies in between the rank of the test message and its own. If a rank violation is observed, the test message is discarded and the sub-DODAG gets either disconnected or a local repair is started (see Fig. \ref{fig:trail_single_validation}). The test message eventually arrives at the root, which adds the current version number to the test message and signs for its way back to the initiating client. Before forwarding, every node verifies whether the signed message contains the scribed rank $j$ that is larger than its own rank. A violation stops the propagation of the message. On reception, the client verifies the signature, matches its nonce, and obtains evidence of the current version number and the rank advertised by its parent. As the rank announcement had consistently travelled to the root, no honest node on the path had observed a rank violation and the upstream is valid. A child not receiving the reply, continues without positive attestation of its parent. It may choose another upstream, if available, or apply additional measures for transport security.

After all nodes have applied this test recursively down the hierarchy with success, it is assured that none of the nodes has a parent that illegally lowered its rank. The highest ranked node that unsuccessfully performs the test identifies the root of the largest sub-DODAG affected by rank spoofing. It should be noted, though, that a directly connected chain of $k$ malicious nodes can secretly replay rank values $k-1$ times so that they are counted in the test as one node. However, this costly attack does not decrease rank values of the attackers, but solely extends the wireless reach of the malicious group and cannot be observed without surveillance of the wireless geometry.

As every node in the network needs to inquire with the root individually, the overhead in messages and signature processing grows linearly with the network size. Hence, this simple scheme of path validation suffers the obvious drawback of scalability. In the following, we will present an aggregated scheme that keeps messages per node and signature computation constant.
%Even though reliable from first hand principles, 

\subsection{Scalable Path Validation}

\subsubsection{Rank Attestation Scheme}
The path validation can be turned into a scalable procedure by aggregating all client-specific inquiries into a single message exchange. Starting from the leaf nodes of a DODAG, we design a convergecast that reaches up to the root. The root node receives and signs a single, converged request that serves as a universal path attestation message when distributed downtree via multicast. 

%\begin{wrapfigure}{l}{0.42\columnwidth}
% 	%\centering
% %	\twocolumn
% 	\includegraphics[width=0.45\columnwidth]{02_TRAIL_DODAG_filter_merge}
% 	\caption{\textsc{TRAIL Rank Verification}}
% 	\label{fig:trial_filter_merge}
%\end{wrapfigure}
%\par

After a leaf node $N_{l,k}$ of the DODAG has received the rank advertisement of its parent (and discovered that it has no further children), it issues a nonce $\eta_{l,k}$ to its parent. The parent node collects the nonces  $\{\eta_{l,k}\}_k$ of all children and writes them into a single array element. For space efficiency, the parent combines the nonces in a Bloom filter \cite{b-stthc-70}. Note that this Bloom filter can be very short, as the number of entries is limited by the number of children per node. This array element containing a single Bloom filter is sent upstream to the grandparent and saved by the node.

From each of its children, the grandparent receives such an array of Bloom filters together with an individual nonce. It should be noted that these arrays need not be of equal lengths, as the tree may be unbalanced. The grandparent aligns every array on the position below the child node rank and merges the entries of equal index using the scalable Bloom filter technique of Almeida et al. \cite{abph-sbf-07}. In detail, the grandparent node extracts all first index elements $A_i(1)$, merges them and writes the result to a new output array $B$ at the index $2$ (incremented by one). In general, $\{A_i(k)\}_i$ are merged into $B(k+1)$, if existent. Finally, the node adds the Bloom filter that aggregates all nonces of its immediate children to the array element $B(1)$ forwards the array $B$ upwards together with its own nonce and saves both $B$ and its nonce. 
%\begin{wrapfigure}{l}{0.42\columnwidth}
 	%\centering
 %	\twocolumn
\begin{figure}
 	\centering
 	\includegraphics[width=0.7\columnwidth]{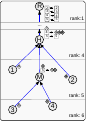}
 	\caption{\textsc{TRAIL Rank Verification}}
 	\label{fig:trial_filter_merge}
\end{figure}
%\end{wrapfigure}
%\par

As depicted in Fig.~\ref{fig:trial_filter_merge}, in proceeding this way stepwise towards the root, an array is created whose index represents the rank and whose values are merged Bloom filters of all nonces issued at a specific rank. Thereby array elements are of variable length, each accommodating the concatenated Bloom filters as generated according to the shape of the tree. Additionally every node on the path saves the array and nonce they forward for latter validation. The root node adds the current version number and signs the data structure consisting of the Bloom filter array and the version number. Thereafter, the signed data is distributed via multicast down the tree. 

\begin{figure}
\begin{center}
	\subfigure[Detection of Duplicate Nonces]{\includegraphics[width=0.23\textwidth]{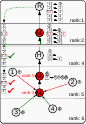} \label{fig:trail_filter_merge_copy}} 
	\subfigure[Detection of Removed Nonces]{\includegraphics[width=0.23\textwidth]{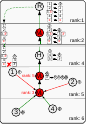} \label{fig:trail_filter_merge_remove}}
	\caption[Nonce Duplicate and Removal Detection]{\textsc{Nonce Duplicate and Removal Detection -- } Squares denote array elements, circles denote single nonces. Attacker $M_1$ copies the nonces from the correct array element to the index of the spoofed rank of $M_3$. In a) Nodes check for duplicates and detect the modification. In b) $M_1$ removes the duplicate nonces. Honest node $H$ detects missing nonces and drops the message.} 
    \label{fig:trail_filter_merge_copy_remove}
        \end{center}
\end{figure}

On the reception, each node can verify the version, and the rank of its parent. It accesses the corresponding array element to match its nonce in the Bloom filter and verifies that no further array element contains the same nonce. Finally, it verifies that the signed Bloom filter array does not contain less nonces than the previously saved array. Note that the probability of a false positive hit can be chosen sufficiently low when configuring the Bloom filter. A successful match testate that ranks have increased monotonically from the root downwards and that the array and contained nonces have not been manipulated or reordered. Whenever the matching fails (see Fig. \ref{fig:trail_filter_merge_copy_remove}), monotonic rank order has been violated on the upward path from the current node to the root. The highest ranked node detecting such violation forms the root of an inconsistently connected sub-DODAG. Any node experiencing such inconsistency may choose another upstream, if available, or apply additional measures for transport security. 
%It should be noted that a directly connected chain of $k$ malicious nodes cannot be detected when secretly replaying rank values $k-1$ times without recording in the test array. They are seen as one node in this test. 

\subsubsection{Security Proof}

We show that a malicious node cannot improve its rank by modifying the data structure, and that improper modifications are detected in the verification phase.

%
%{\em Assumptions:} We rely on the attacker model specified in Section \ref{sec:attacker_model}. In particular, we refer to a single attacker that has no collaborators, and no means to manipulate radio transmission channels or out-of-band communication. As has been discussed above,  malicious neighbors in a topological chain will be able to act under a common rank without discovery of any of the security schemes discussed in this paper. Distributed attackers scattered among different hierarchy levels that can coordinate out-of-band will also be able to undermine the attestation procedures of TRAIL, VeRA etc.. We will also exclude effects of false positives inherited from Bloom filters, as these probabilities can be made arbitrarily small, and remain without effect in normal operations. Security threats only arise at the coincidence of a false positive and an attack to the corresponding element.

\smallskip
\noindent
{\em Assumptions.} We rely on the attacker model specified in Section \ref{sec:attacker_model}. In particular, we refer to an attacker that has no means to establish an out-of-band communication channel. A chain of $k$ malicious neighbors is considered as one attacker with an extended wireless reach. Distributed attackers scattered among different hierarchy levels communicating out-of-band channel cannot be detected and are not considered in our model. However, non-collaborating attackers distributed in the topology are considered. Finally, we ignore the false positive rates on queries to bloom filters as they can be made arbitrarily small by choosing appropriate parameters. 
%We will also exclude effects of false positives inherited from Bloom filters, as these probabilities can be made arbitrarily small, and remain without effect in normal operations. Security threats only arise at the coincidence of a false positive and an attack to the corresponding element. 
%It is possible for the attacker to have multiple collaborating nodes placed in the topology. The collaboration is limited to the knowledge of their topological positions. In addition the nodes cannot communicate in-band which each other.

%{\em Proof:} 
\smallskip
\begin{IEEEproof}
We consider the security of TRAIL in existence of i) multiple non-collaborating malicious nodes and ii) multiple malicious nodes with limited collaboration:

{\em i) Multiple non-collaborating malicious nodes:} Since the nodes are not allowed to collaborate, they can be considered as multiple single attackers. For simplicity, we provide the analysis for a single malicious node. A malicious node receiving a topology test message  $\langle \eta, A \rangle$ from its child(ren) has the option to (1) not include its child(ren) in the message array or to not merge-and-forward the array $A$ at all. It may as well  (2) rearrange the array, and in particular include the nonces of its child(ren) at a wrong array position. It may (3) attempt to exclude itself from the attestation hierarchy by not submitting its nonce value to its parent. These four choices of malicious nodes will lead to the following conditions:
%Finally collaborating attacker (4) may rearrange the array to improve ranks of distant accomplices. These four choices of malicious nodes will lead to the following conditions:
%\begin{description}
%\item[(1):]
\begin{enumerate}
  \renewcommand*\labelenumi{C\theenumi.}
  \item
By not forwarding the test nonces of its children or the attestation array, the malicious node causes its immediate detection. When receiving the signed attestation message of the root, the child(ren) of the malicious node will test for its nonces without success and detect the inconsistency. 
  \item
The best a malicious node can do to its children is writing nonces at the foreseen position. Any misplacement will move data of the children to a lower rank position and thus cannot be aligned with a malicious rank upgrade. Other rearrangements of the array will change the data positions for nodes lower in the tree. This implies that affected nodes are not within the wireless transmission range of the malicious node -- they had chosen the better rank of the malicious node  otherwise. As the malicious node cannot coordinate rank advertisements outside its wireless reach, nodes will remain unaware of their nonce moving to other rank positions. Nodes will thus search at the original rank position in the attestation message and corresponding tests will fail. 
  \item
If the malicious node withholds its own nonce, but cooperates in traversing the merged filter array, its honest parent will merge the data with data from its other children and insert at the proper position. Not delivering the nonce will simply lead to a Bloom filter that does not contain the nonce of the malicious node. Hence, an malicious node causes nothing but excluding itself from the verification process.  
\end{enumerate}

\smallskip
{\em ii) Multiple malicious nodes with limited collaboration:} We mean by a limited collaboration that multiple attackers know in advance their position in the topology and the desired rank which they want to claim during an attack. This can be realized by configuring them accordingly during their deployment. Limited indicates that once they are deployed, those malicious nodes, which are not within each other's communication range, cannot communicate anymore. TRAIL mitigates such attacks as follows: 
%A collaborating malicious node close to the root may merge Bloom filter elements from its lower positioned accomplice and its sub-DODAG. 
A malicious node close to the root merges array elements on behalf collaborating malicious nodes lower in the topology that claim a false rank. %ranks to the position of ranks claimed by collaborating nodes lower in the topology. 
Consequently, nonces of honest nodes that are affected by the rank spoofing, are moved to the correct array element. However, due to the malicious merging of array elements, these nonces exist multiple times. Such a duplicate either denotes a fraud or a false positive. Given a false positive rate of $f$, we detect the attack with probability $1-f$. Deleting nonces from filters will cause that an honest node on the path will detect the attack by comparing the forwarded array with the signed one.

In any of the cases, forgery will not comply to a rank decrease and will be detected, whenever it affects third party nodes. All parents of a malicious node will always exclusively write to the lower rank-test positions, which is the  obvious protection from rank spoofing in this procedure.
\end{IEEEproof}

\subsubsection{Details of the Bloom Filter}
 
We use Bloom filters \cite{b-stthc-70}, a space-efficient random data structure, to reduce message lengths in our attestation scheme. 
A Bloom filter is defined as a bit-vector, $v$ of $m$ bit and represents a data set. By using $k$ independent hash functions, each element of a set of $A=\{a_{1}, \dots , a_{n} \}$ is mapped to $k$ bits in $v$. By these means, the size of each input element is reduced to at most $k$ bits. Due to randomized overlapping of bits from different elements, the size may be reduced even further, but this may return a false positive result of a query. Essentially, there is a linear relation between number of bits used for storing each element, and the false positive rate. Mitzenmacher \cite{m-cbf-02} could show that properly designed Bloom filters can be compressed even further by about 30 \% at a given false positive rate. Almeida et. al. \cite{abph-sbf-07} designed a scalable extension of Bloom filters that linearly add filter elements with increasing set sizes. 

In TRAIL, we require tiny Bloom filters that store nonces from the children set of a single node. For a commonly small fanout of $k$ nodes and a false positive rate below $1 \%$, an appropriate bit-size $m$ of the (compressed) Bloom filters can be estimated as $m = 6\, k $ [bits].

\section{Performance Evaluation}
\label{sec:eval}

\begin{figure}
	\centering
	\includegraphics[width=1.0\columnwidth]{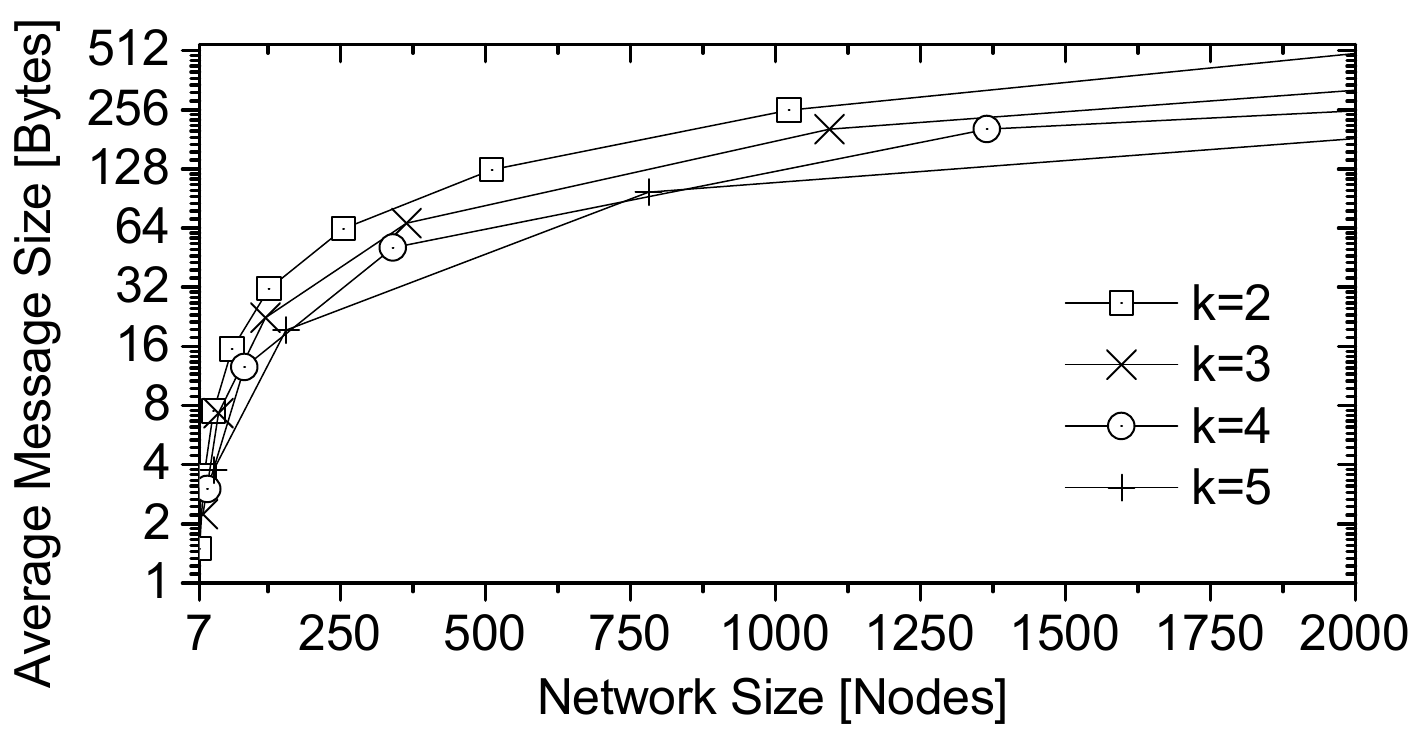}
	\caption{\textsc{TRAIL Message Sizes}: Average message size distribution for varying fanout degrees $k$ as functions of the network size.}
	\label{fig:message_size}
\end{figure}

% ideally the TRAIL starts at leafs and propagates upwards: timer as in RPL
% each message must arrive: ACK

% Complexity
%	- calulation each node O(k) Bloom filter
%	- transmission each node O(N+1) where N amount of child nodes
%

To evaluate the performance of our RPL security scheme, we have implemented  TRAIL authentication (attestation and announcement messages)  as an extension to the existing RPL protocol implementation on the RIOT platform \cite{bhgws-rotoi-13}---an integration of TRAIL in RPL ICMP control messages has little effect on performance and is left to future protocol engineering work. We deployed TRAIL on the DES Mesh Testbed of FU Berlin \cite{gbj-cdhde-08} and performed the comparative experiments described below. The focus of our evaluation lies on the overhead cost  and the temporal performance of TRAIL in comparison to unmodified RPL routing. The critical cost metrics for wireless sensor nodes are over the air transmission, e.g., the number of messages sent, as well as message sizes. 

First we analyze the message characteristics of TRAIL as a function of the network size. The  critical resource consumption of TRAIL is given by the sizes of the attestation messages. As nodes need to accumulate nonce values of their parent nodes, the attestation array grows with increasing network sizes. While messages are tiny at the leaf nodes, the array gets larger towards the root node. Fig.~\ref{fig:message_size} visualizes the average message sizes  for different fanout degrees $k$ of the inner nodes as functions of the total network size. For simplicity, we assume balanced $k$-ary trees, but results are not strongly dependent on tree shapes.  
It is clearly visible that small message sizes compliant to 6LowPAN MTUs constrain network dimensions by about $\approx 250$ nodes. The characteristic performance aspects of TRAIL for different network sizes and tree configurations are summarized in Table~\ref{tbl:trail-msgoverhead}, from which we can extract the extra traffic imposed by TRAIL: Two messages per node at the given size distribution. 

\setlength{\fcwidth}{0.15cm}
\begin{table}

  \caption{Message overhead for different network sizes: (\textnormal{$k$=
  number of children, $h$=height of the tree})}\label{tbl:trail-msgoverhead}
  \begin{tabular}{llp{0.85cm}p{1.3cm}p{1.0cm}p{1.0cm}}
   \toprule
    \multicolumn{3}{c}{Network Configuration} & \multicolumn{3}{c}{Message
    Overhead [Bytes]}\\
    \cmidrule(lr){1-3} \cmidrule(lr){4-6}
    $k$ & $h$ & \# Nodes & \# Msg. per node & Average Size & Max. Size
    \\ \midrule

    \multirow{3}{\fcwidth}{2} & 3 & 15 & 2 & 3.5 & 10.5\\
     & 4 & 31 & 2 & 7.5 & 22.5 \\
     & 5 & 63 & 2 & 15.5 & 46.5 \\

    \midrule

     \multirow{3}{\fcwidth}{4} & 3 & 85 & 2 & 12.6 & 63 \\
     & 4 & 341 & 2 & 51 & 255 \\
     & 5 & 1365 & 2 & 204.6 & 1023\\
   \bottomrule
  \end{tabular}
 % \vspace{-5mm}
\end{table}

\begin{figure*}
	\centering
	\subfigure[\textsc{Experimental DODAG with Attacker}]{\includegraphics[width=0.75\columnwidth]{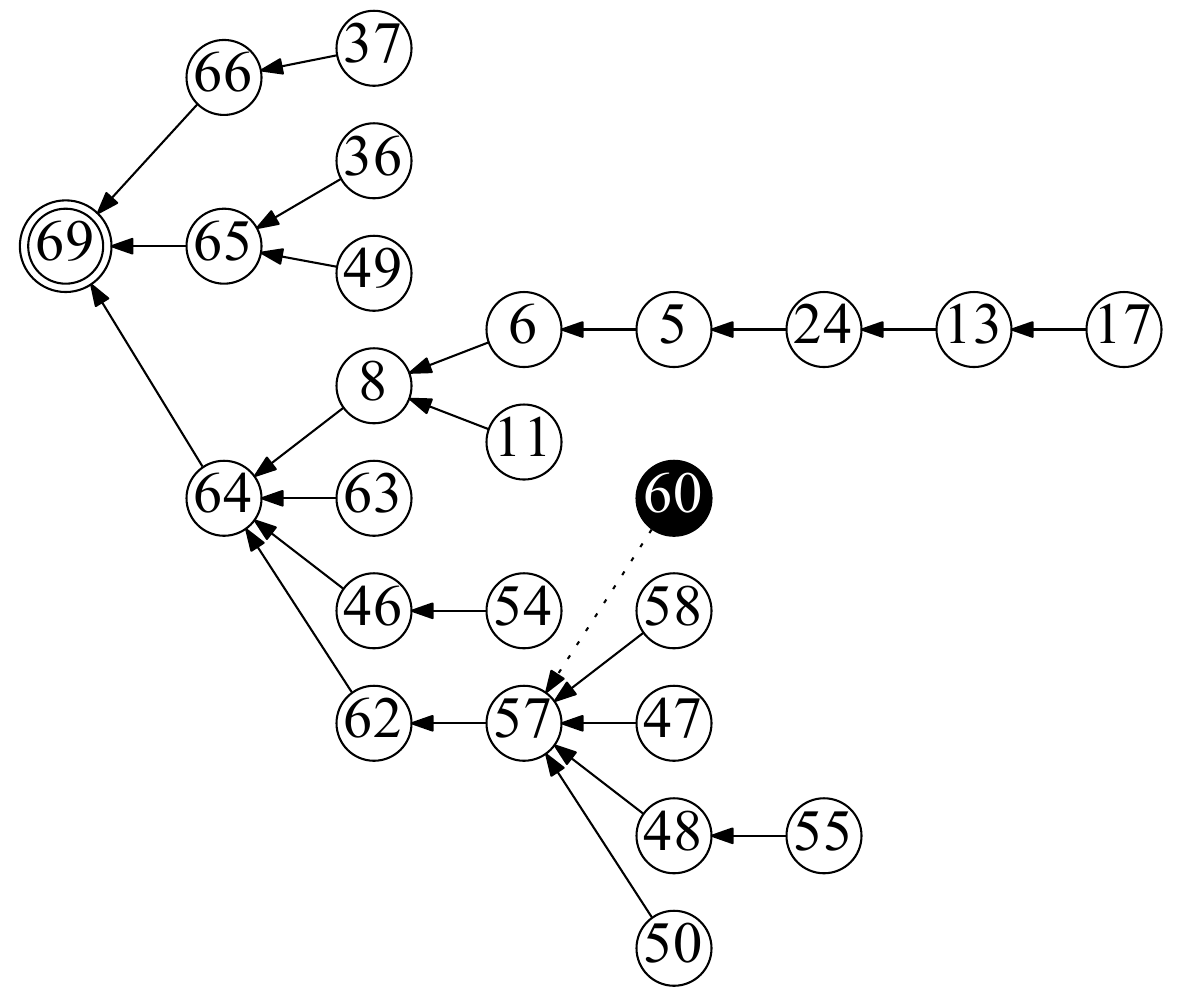}\label{fig:experiment}}\qquad
	\subfigure[\textsc{Routing Convergence Times in Rank Order}]{\includegraphics[width=1.2\columnwidth]{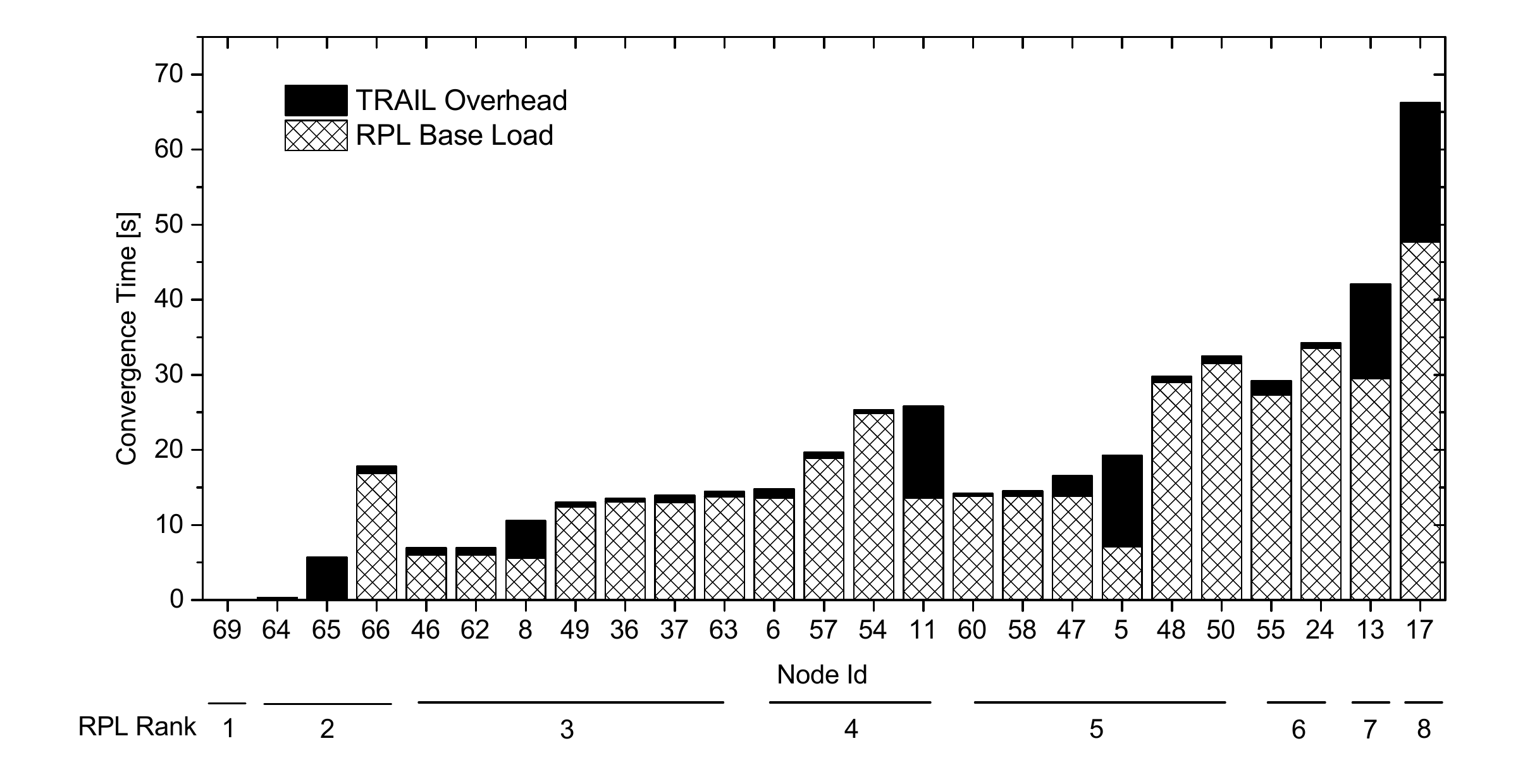}\label{fig:convergence}}
	\centering\caption{Per node performance in joining the DODAG: Pure RPL and the TRAIL overhead as observed in the testbed.}
	
\end{figure*}

Our second evaluation targets at the temporal performance of route convergence. We deployed TRAIL on 25 MSBA2 nodes distributed in the sensor network testbed and compared with an identical pure RPL installation. RPL/TRAIL arranged a DODAG with the highest rank of eight as visualized in Fig. \ref{fig:experiment}. Choosing an attacker (node 60) to announce a root rank led to a break up of the pure RPL network. Only seven nodes remained in the initial, upright network, while 17 nodes reconnected to the bogus tree. TRAIL discovers  and isolates the attacker immediately with its (bogus) rank announcement. As a consequence, TRAIL rearranged a connected tree of all honest neighbors, excluding the bogus node. 

Routing convergence times for tree construction were measured node-wise during the experiments. Comparisons between the pure RPL and the overhead induced by TRAIL are plotted in Fig. \ref{fig:convergence}. Naturally, the wireless ad-hoc regime produces large variations that are visible for both, RPL and TRAIL. Nevertheless, the additional times needed to join a DODAG while performing the security extensions of TRAIL remain below 20 \% in most cases. Occasional authentication messages exceed this limit due to message loss and retransmissions. However, such performance fluctuations are characteristic for all mesh routing operations including RPL.

%\vspace*{-2mm}

\section{Conclusions \& Outlook}
\label{sec:conclusion}

This work focuses on routing security of RPL, a recent routing protocol for the emerging Internet of Things. Intrinsically, RPL is vulnerable to topology attacks. Its rank and version number need particular protection, since by spoofing version and rank an attacker can obtain dominant impact on the network. The current state of the art leaves relevant security issues unresolved. 

Our first contribution in this paper was to analyze and improve VeRA, a cryptographically centered  protection scheme. We identified new attack vectors and modified VeRA to withstand them. While returning to the topological core of the problem, our second contribution introduces TRAIL. TRAIL defines a test procedure to inquire on the actual path properties of the routing system. This generic approach is built on first-hand principles and -- different from VeRA -- requires almost no cryptography. Its main cryptographic workload is carried out by the root node, which acts as a (stronger) gateway in typical RPL deployments.
 TRAIL is designed to minimize network message exchanges and node resource consumption. Our evaluations revealed that the transmissions of bits required by TRAIL remain feasible for typical challenged environments, and that a testbed of typical shape can well operate TRAIL with limited additional effort. 
Future directions of this work are twofold. First, we will further optimize our algorithms to reduce dependency on network sizes. 
Second, we intend to apply the TRAIL approach proposed for RPL to other routing protocols. 
%Second, we intend to take advantage of the generality of the TRAIL approach to develop a generic topology monitoring that likewise works for other routing regimes. 

%\hfill SAFEST
%\hfill \today%January 11, 2007
%\subsection{Subsection Heading Here}
%Subsection text here.
%\subsubsection{Subsubsection Heading Here}
%Subsubsection text here.

%\section*{Acknowledgments}
%We thank Emmanuel Baccelli for fruitful discussions about RPL. This work is partly supported by the German BMBF within the projects SAFEST
%and SKIMS.
%This work is supported by the BMBF project SAFEST -- \textit{Social-Area Framework for Early Security Triggers at Airports}

%\pagebreak

%\nocite{*}
\balance
\bibliographystyle{IEEEtran}
\bibliography{rfcs,ids,security,own,manet,theory}
%\bibliography{Infocom2013_SAFEST}

\end{document}